# Dual-Gate Modulation of Carrier Density and Disorder in an Oxide Two-Dimensional Electron System

Zhuoyu Chen[1*], Hongtao Yuan[1,2], Yanwu Xie[1,2], Di Lu[1], Hisashi Inoue[1], Yasuyuki Hikita[2], Christopher Bell[2,3], Harold Y. Hwang[1,2*]

[1]*Geballe Laboratory for Advanced Materials, Stanford University, Stanford, CA 94305, USA.*

[2]*Stanford Institute for Materials and Energy Sciences, SLAC National Accelerator Laboratory, Menlo Park, CA 94025, USA.*

[3]*H.H. Wills Physics Laboratory, University of Bristol, Tyndall Avenue, Bristol, BS8 1TL,UK.*

*Corresponding authors: zychen@stanford.edu, hyhwang@stanford.edu

**Abstract** Carrier density and disorder are two crucial parameters that control the properties of correlated two-dimensional electron systems. In order to disentangle their individual contributions to quantum phenomena, independent tuning of these two parameters is required. Here, by utilizing a hybrid liquid/solid electric dual-gate geometry acting on the conducting $LaAlO_3/SrTiO_3$ heterointerface, we obtain an additional degree of freedom to strongly modify the electron confinement profile and thus the strength of interfacial scattering, independent from the carrier density. A dual-gate controlled nonlinear Hall effect is a direct manifestation of this profile, which can be quantitatively understood by a Poisson-Schrödinger subband model. In particular, the large nonlinear dielectric response of $SrTiO_3$ enables a very wide range of tunable density and disorder, far beyond that for conventional semiconductors. Our study provides a broad framework for understanding various reported phenomena at the $LaAlO_3/SrTiO_3$ interface.

Various quantum phenomena in two-dimensional (2D) electron systems are controlled by carrier density and disorder. For instance, 2D superconductor-insulator transitions can be induced both by tuning the carrier density (e.g. in gated cuprate thin films[1,2]), and by varying the disorder level (e.g. in amorphous metal thin films[3,4]). However, these two parameters are entangled experimentally in correlated systems like oxides: whether by chemical or electrostatic doping, varying the carrier density is usually accompanied with a change in the effective disorder as probed by mobility, such as the change of interface scattering with gate voltage in field-effect devices[5,6]. This hinders the understanding of individual contributions from the two parameters to 2D quantum phenomena, especially when interactions between electrons are prominent, such as modulating the superconducting transition temperature $T_c$ in transition metal oxides[6-8]. Therefore, an experimental method to effectively and independently tailor carrier density and disorder in 2D systems would be ideal. In principle, this can be achieved for the interface confined 2D electron gas in semiconductors, via dual electrostatic gate controlled deformation of the electron envelope wavefunction[9].

The LaAlO$_3$/SrTiO$_3$ (LAO/STO) heterointerface[10] serves as a prototypical system for this purpose since it not only possesses interface induced symmetry breaking[11], but also hosts a range of intriguing emergent phenomena including tunable 2D superconductivity[6,7,12] and magnetism[13-15] within a high-mobility system[16,17]. Towards understanding these fascinating properties coherently, a major difficulty lies in the incorporation of the large and nonlinear dielectric response of STO (ref. 18,19) as well as interface disorder into the paradigm of confined electrons with $d$-orbital effective mass anisotropy[20-22]. In order to address this problem, the nonlinear Hall effect, which is induced by the existence of parallel conduction channels with different mobilities, can be a useful probe of the effect of anisotropic effective mass and disorder. It has been found that the modulation of the nonlinear Hall effect has an apparent correlation with the superconducting dome in STO heterointerfaces[23,24], which also suggests a potential avenue for understanding superconductivity at this interface. Importantly, the large dielectric constant of STO greatly enhances gate control of the quantum well profile for disorder modulation far beyond that for conventional semiconductors[9]. Here we demonstrate a hybrid ionic liquid/solid dual gate technique in the LAO/STO system probing the interface properties over a broad range of carrier density and mobility. This dual-gate approach allows effective and independent tuning of carrier density and disorder within a single device. The observed systematic dual-gate tuned nonlinear Hall effect shows that disorder modulation by gating dominates the change in mobility and the evolution of the Hall nonlinearity, which is highly consistent with Poisson-Schrödinger calculations of the dual-gate tuned interface confinement.

The concept of our dual-gate devices is to control the interface and bulk boundary conditions of the asymmetric quantum well via the top[25] and back[6,26] gates, respectively[9,27]. As sketched in Figure 1a, with decreasing top gate voltage ($V_{TG}$) at fixed zero back gate voltage ($V_{BG}$), the electric field at the interface becomes weaker and the electron envelope wavefunction is pushed *away* from the interface and its associated scattering, resulting in an increase of the mobility. Conversely, with decreasing $V_{BG}$ at fixed $V_{TG}$ (Figure 1b), the electric field in the quantum well, especially on the STO bulk side, becomes stronger and the electron wavefunction is compressed *towards* the interface, thus the mobility tends to decrease. When the two gates work simultaneously, for example, by increasing $V_{TG}$ and decreasing $V_{BG}$ at the same time, a fixed carrier density and a varied disorder level (i.e. mobility) can be obtained.

Our dual-gate devices are formed by simultaneously gating from the top of the epitaxial LAO layer and the back of the STO substrate. As shown in Figure 1c, the top gate is implemented using an ionic liquid (IL), which has been proven exceptionally effective for tuning electron density and interfacial electric field in various materials over broad ranges[28–36]. In contrast to previous studies where the IL is in direct contact with the conducting channel, here we apply the IL (DEME-TFSI) on top of the ultrathin epitaxial LAO grown by pulsed laser deposition, which serves as a separating layer between the IL and STO. The LAO layer protects against electrochemical reactions in the biased STO, and thus stabilizes the electrostatic modulation of the interface. This can be seen from a very small hysteresis in the top gate transfer curve as shown in Figure 1d, as compared to cases with strong electrochemical effects[37,38]. A broad modulation range is achieved even with the inserted LAO layer due to the 4 unit cell (1.5 nm) thickness. Further details are provided in Supporting Information (SI). For both the top ionic liquid gate and the back solid gate, the LAO/STO interface channel can be turned ON and OFF, as shown in Figure 1d and 1e.

We first focus on the top gate tuning with $V_{BG}$ fixed to zero. Figure 2a shows the Hall resistance $R_{xy}$ as a function of the perpendicularly applied magnetic flux density $B$ under various $V_{TG}$ at temperature $T = 2$ K. The large change in $R_{xy}$ for different $V_{TG}$ indicates a wide range of carrier density modulation. In addition, the Hall effect is largely linear at lower $V_{TG}$, whereas at higher $V_{TG}$, the Hall effect becomes significantly nonlinear (magnified in inset). Figure 2b shows the Hall density extracted from the slope $dR_{xy}/dB$ at 13 T ($n_{total} = (e\ dR_{xy}/dB|_{B=13T})^{-1}$, where $e$ is the electron charge) and in the low field limit ($n_{0T} = (e\ dR_{xy}/dB|_{B=0T})^{-1}$). The high field data $n_{total}$ captures most of the mobile electrons at the interface because the slope is saturating at 13 T, and thus $n_{total}$ estimates the total carrier density. Further discussions are provided in SI. The $n_{total}$ data show that the top gate is able to tune the density linearly from $1 \times 10^{13}$ cm$^{-2}$ to higher than $1 \times 10^{14}$ cm$^{-2}$. When comparing $n_{total}$ and $n_{0T}$, several notable features are seen: in the low-$V_{TG}$ regime, $n_{total}$ overlaps with $n_{0T}$, indicating a linear Hall effect; however, when $V_{TG}$ is high, $n_{total}$ becomes much larger than $n_{0T}$, in accordance with the substantial Hall nonlinearity in this regime. A crossover from linear to nonlinear Hall effect can be seen at around $V_{TG} = 1$ V. The ratio $n_{0T}/n_{total}$ quantifies the strength of the Hall nonlinearity: smaller $n_{0T}/n_{total}$ corresponds to stronger Hall nonlinearity, as indicated by the background colour scale.

The inset of Figure 2b plots the temperature dependence of the Hall densities in different $V_{TG}$ regimes. In the low-$V_{TG}$ case, the Hall effect is principally linear ($n_{total} \simeq n_{0T}$) at all temperatures. In the high-$V_{TG}$ case, however, the Hall nonlinearity has an obvious temperature dependence: the nonlinearity is negligible ($n_{total} \simeq n_{0T}$) at 150 K and above, but becomes prominent ($n_{total} > n_{0T}$) during cooling at around 100 K (ref. 39-41). When the temperature is lower than 10 K, the discrepancy between $n_{total}$ and $n_{0T}$ saturates. 100 K and 10 K match the temperatures where the STO relative dielectric response $\varepsilon_r$ starts to rise and saturate, respectively[18,19], suggesting that the nonlinearity in the Hall effect is related to the large and nonlinear dielectric response of STO.

The nonlinear Hall effect can be well fitted using the two-carrier model, and results are shown in Figures 2c and 2d (for fitting procedures see SI). The crossover at $V_{TG} = 1$ V is even clearer when examining the two carrier decomposition. In the low-$V_{TG}$ regime, essentially all carriers are high-mobility carriers $n_{HM}$ and their mobility $\mu_{HM}$ decreases with the increase of $V_{TG}$. In contrast, for $V_{TG} > 1$ V, the low-mobility carrier density $n_{LM}$ (= $n_{total} - n_{HM}$) increases dramatically and dominates over $n_{HM}$, while the corresponding mobility $\mu_{LM}$ monotonically decreases. Meanwhile, $n_{HM}$ and $\mu_{HM}$ remains basically constant. The total effective carrier mobility $\mu_{total} = \sigma/en_{total}$, where $\sigma$ is the channel sheet conductivity, is shown in Figure 2d. For all $V_{TG}$, $\mu_{total}$ decreases with increasing $V_{TG}$ (ref. 42).

Back gate tuning data are shown in Figure 3. $V_{TG}$ is fixed at different values in these measurements. As shown in Figure 3a, in a typical low-$V_{TG}$ case ($V_{TG}$ = 0.9 V), the Hall effect displays linear behavior ($n_{13T} \simeq n_{0T}$) for all $V_{BG}$. By contrast, in the high-$V_{TG}$ regime ($V_{TG}$ = 2.5 V), strong nonlinearity occurs ($n_{13T} > n_{0T}$), but it can be suppressed by decreasing $V_{BG}$ ($n_{13T}$ approaches $n_{0T}$, also see inset for $R_{xy}(B)$ data)[6,23,24]. Note that this suppression of nonlinearity is totally different from the top gate case, since here the loss of nonlinearity in Hall effect occurs at a much higher density around $6 \times 10^{13}$ cm$^{-2}$. Interestingly, the two-carrier fitting results (Figure 3b) show that the back gate suppression of the Hall nonlinearity is mostly coming from the decline of $\mu_{HM}$ with decreasing $V_{BG}$, whereas $\mu_{LM}$ is basically unaffected. Another key point is that unlike the decline of $\mu_{total}$ with $V_{TG}$, $\mu_{total}$ continuously increases with the increase of $V_{BG}$ (ref. 6).

When we combine the top and back gates, a two-dimensional parameter space spanned by carrier density and mobility is obtained[43], as shown in Figure 4, in which each point corresponds to a ± 13 T field sweep (i.e. a measure of the total carrier density $n_{total} = n_{LM} + n_{HM}$, independent of distribution). Both mobility and density are tuned by more than an order of magnitude in the dual-gate devices. Importantly, we are able to keep density unchanged and investigate the effect from mobility only, and vice versa, by the combined manipulation from both gates. By assuming a simple Drude model, in which $\mu_{total} = e\tau/m^*$ ($\tau$ is the average scattering time, and $m^*$ is the average in-plane effective mass among different $t_{2g}$ orbitals[21,44]), we are able to extract $\tau$ from $\mu_{total}$, as a more direct measure of the effective disorder level of the system shown on the right axis of Figure 4. In this wide 2D parameter space, the dual-gate modulation of the nonlinear Hall effect can be summarized: the nonlinear Hall effect can be suppressed by reducing either (1) the low-mobility electrons with decreasing $V_{TG}$ or (2) the high-mobility electrons with decreasing $V_{BG}$. The background colour scale in this figure represents the STO $\varepsilon_r$ at the interface estimated based on Poisson-Schrödinger simulations, as discussed below. It clearly shows the strong correlation between the emergence of low-mobility electrons and the collapse of the dielectric constant at the interface.

In order to understand the experimental results, particularly the nonlinear Hall effect in the dual-gate structure, we performed a non-uniform-meshed self-consistent Poisson-Schrödinger calculation[45] incorporating the large and nonlinear electric field dependent dielectric constant $\varepsilon_r$ of STO arising from its incipient ferroelectricity[18,19,46]. Whereas STO possesses a very large $\varepsilon_r$ (~ $10^4$) when the electric field is small at a temperature lower than 10 K, $\varepsilon_r$ drops dramatically when the polarization is saturated at high electric fields. In addition to considering the anisotropic effective mass for the $t_{2g}$ bands, Fermi-Dirac statistics are also included in the model, and the temperature is set to 2 K in the following results. The top and the back gate voltages are estimated by extrapolating the electric potential at the boundaries of the calculation to the positions for the gate electrodes with reference to the Fermi level (for more details of the calculations, see SI). The calculation results show that all the above experimental observations are highly consistent with a picture of dominant disorder modulation, as opposed to the effective mass anisotropy of the $t_{2g}$ subbands, as we discuss in detail below.

Figure 5a and 5b show the quantum well profile and the electron distribution in the depth direction (z), respectively, under varying $V_{TG}$ and fixed $V_{BG} = 0$ V (top gate mode). Note that here we plot the locally averaged effective potential in a continuum approximation. At low $V_{TG}$ (< 1.0 V), the electric field at the interface is small so that the large $\varepsilon_r$ leads to relatively weak confinement. Electrons are mostly distributed in a wide region deep ($\gtrsim$ 5 nm) into the STO substrate (corresponding to high-mobility electrons $n_{HM}$ in Figure 2c), and there are only few electrons located near the more disordered interface (these represent the low-mobility electrons $n_{LM}$). Therefore, the Hall effect measurement in this regime is principally linear. When $V_{TG}$ is increased in this regime, the confinement increases so both $\mu_{HM}$ and $\mu_{LM}$ decrease (Figure 2d).

$V_{TG}$ = 1 V is the crossover point where the interface $\varepsilon_r(z = 0)$ starts to collapse, so when $V_{TG}$ > 1 V, $\varepsilon_r(z = 0)$ drops to modest values. Under this condition, due to the small value of $\varepsilon_r(z = 0)$, the confinement potential at $z = 0$ is relatively steep, leading to an accumulation of significant electron density very near the disordered interface. These electrons are strongly scattered by interface disorder, and thus have lower mobility, corresponding to $n_{LM}$ in Figure 2c. The more $V_{TG}$ increases in this regime, the larger is the density of these confined electrons (thus $n_{LM}$ is enhanced), and the confinement increases (thus $\mu_{LM}$ decreases, seen in Figure 2d). In contrast to these strongly confined electrons, a weakly confined population of electrons still exists (electrons distributed at $z \gtrsim$ 5 nm, corresponding to $n_{HM}$), because in this region $\varepsilon_r$ has restored its low-field value. Note that the spatial distribution of these weakly confined electrons does not change with $V_{TG}$. This explains why in the regime when $V_{TG}$ > 1 V, both $n_{HM}$ and $\mu_{HM}$ remain unchanged with $V_{TG}$ (Figure 2c). Due to the existence of both populations of electrons and their distinctly different mobility, the Hall effect in this regime is highly nonlinear. The temperature dependence of the nonlinear Hall effect (Figure 2b inset) is also consistent with the simulation. When the temperature is higher than 100 K, since the $\varepsilon_r$ of STO is relatively small at all electric fields[18,19], all electrons are strongly confined near the interface, leading to a linear Hall effect. By contrast, at temperatures lower than 100 K, the $\varepsilon_r$ of STO becomes large and strongly electric field dependent, resulting in the emergence of different confinement strength of electrons and thus the strong nonlinearity in Hall effect.

Figure 5c and 5d show the quantum well profile and the electron distribution in the $z$ direction tuned by $V_{BG}$ with fixed $V_{TG}$ = 1.3 V (back gate mode). When the back gate decreases, two important features are found. First, the distribution of the weakly confined electrons is substantially modulated. The decrease of $V_{BG}$ lifts the bulk boundary of the quantum well and thus pushes these electrons towards the interface. This leads to the decrease of mobility for $n_{HM}$ (thus $\mu_{HM}$ decreases in Figure 3b). Second, the distribution of the strongly confined electrons remains basically unchanged under different $V_{BG}$, explaining why $\mu_{LM}$ is largely independent of $V_{BG}$ in Figure 3b. When $V_{BG}$ reaches large negative values, all electrons have been compressed close to the interface, thus effectively leading to a linear Hall effect dominated by $n_{LM}$. The insets in Figure 5a and 5c reveal that the dielectric constant at the interface ($z$ = 0) is strongly dependent on the top gate tuning yet is largely independent of the back gate tuning. This $\varepsilon_r(z = 0)$ is shown in the background colour scale in Figure 4.

According to our calculations, multiple subbands are always filled in the density range probed by the current experiment (See SI for further information on the subband structure). In principle, all subbands are contributing to the Hall effect with their different $t_{2g}$ effective mass and effective disorder level. The two-carrier model is a parameterization that averages over these many subbands and gives an effective grouping of high-mobility and low-mobility electrons. In order to correctly interpret the two-carrier model, a central question to ask is whether the difference in effective mass or the difference in disorder level is dominating the nonlinear Hall effect. In the picture that different effective mass dominates, $n_{HM}$ is assigned to electrons from subbands with small in-plane effective mass (derived from $d_{xy}$ orbitals), and $n_{LM}$ is assigned to electrons from subbands with large in-plane effective mass ($d_{yz}/d_{xz}$ orbitals). Since the effective masses from the different orbitals differ significantly[20], this picture is a reasonable conjecture. Within such a picture, our observation of top gate tuning (Figure 2) could be interpreted as single-type ($d_{xy}$) subband occupancy when $V_{TG} < 1$ V and multiple-type ($d_{xy}$ and $d_{yz}/d_{xz}$) subband occupancy when $V_{TG} > 1$ V. And the sudden saturation of $n_{LM}$ in the $V_{TG} > 1$ V regime would require a negative electronic compressibility. For completeness, a thorough analysis of the negative electronic compressibility scenario is given in the SI[47]. However, the effective mass dominating picture fails to consistently explain many of our other experimental observations. First, the nonlinear Hall effect can be suppressed with decreasing $V_{BG}$ while density is still as high as $6 \times 10^{13}$ cm$^{-2}$, which cannot correspond to single-type-subband occupancy. Second, the modulation of the ratio $\mu_{HM}/\mu_{LM}$ via back gating is over a factor of 3 as shown in Figure 3b, while in the effective mass-dominating picture, the mobility ratio should match the ratio of effective mass and thus should be essentially a constant. Third, recent observation of two frequencies in quantum oscillation experiments performed at low densities ($< 10^{13}$ cm$^{-2}$), where the Hall effect is linear, shows the existence of different effective masses, implying the filling of both $d_{xy}$ and $d_{yz}/d_{xz}$ subbands[17]. Therefore, while we cannot completely rule out any contributions to the electrostatic response arising from correlation effects, our results are most consistently and quantitatively understood as consequences of the large and nonlinear dielectric property of STO and the modulation of effective disorder of the interface electron system.

Based on the above results and discussion, we can frame a broad perspective towards understanding various reported behavior of the confined electrons at the LAO/STO interface[48]. When the density is low, the large dielectric constant of STO gives rise to the weak confinement of the electrons, leading to the advent of high mobility and Shubnikov-de Haas (SdH) oscillations at the interface[16,17,41]. The weak confinement also renders the occupation of multiple subbands in the low density regime, resulting in the two frequencies in the oscillations observed[17]. When the electron density is higher, via either top gating or growth/surface control[41,42,49,50], the dielectric constant of STO at the interface collapses to a smaller high-field value. The resulting emergence of strongly confined electrons leads to the nonlinearity in Hall effect[41] and superconductivity[6,7,12,23,24]. Nevertheless, subbands within the coexisting high-mobility tail can still generate SdH oscillations in high magnetic fields[51]. However, when decreasing the back gate to large negative values, all electrons are compressed against the interface and strongly affected by the interface disorder. The absence of weakly confined electrons leads to the suppression of the nonlinear Hall effect, and the strong disorder induces a superconductor-insulator transition[6,7,24].

In conclusion, we designed and implemented an ionic liquid based dual-gate device using the $LaAlO_3$/$SrTiO_3$ heterointerface. We demonstrate effective manipulation of the disorder level independent of 2D density via modification of the electron distribution, and further spanned a broad (carrier density, mobility) 2D parameter space for the interface electron system. With the capability of the device and the quantitative analysis and modeling, our results construct a consistent understanding of a wide range of reported behaviors of the $LaAlO_3$/$SrTiO_3$ interface electrons, emphasizing the central role of the large and nonlinear dielectric response of STO. This dual-gate device structure can be useful for oxide device applications, as well as probing 2D quantum phenomena in oxide heterostructures.

ASSOCIATED CONTENT

**Supporting Information**. This section provides detailed information about experiment methods, data analysis, and numerical simulations. The following files are available free of charge. (file type: PDF)

AUTHOR INFORMATION

Corresponding Author

* Email: zychen@stanford.edu

* Email: hyhwang@stanford.edu

Notes

The authors declare no competing financial interest.

ACKNOWLEDGMENTS

We gratefully acknowledge Biao Lian, and Satoshi Harashima for their help in experimental and scientific discussions. This work was primarily supported by the Department of Energy, Office of Basic Energy Sciences, Division of Materials Sciences and Engineering, under contract DE-AC02-76SF00515. Partial support was provided by the Stanford Graduate Fellowship in Science and Engineering (Z.C., H.I.), and ONR-MURI number N00014-12-1-0976 (Z.C., Y.X.). Part of this work was performed at the Stanford Nano Shared Facilities (SNSF).

ABBREVIATIONS

LAO, $LaAlO_3$; STO, $SrTiO_3$; IL, ionic liquid; DEME-TFSI, *N,N*-diethyl-*N*-(2-methoxyethyl)-*N*-methylammonium bis(trifluoromethylsulphonyl)-imide.

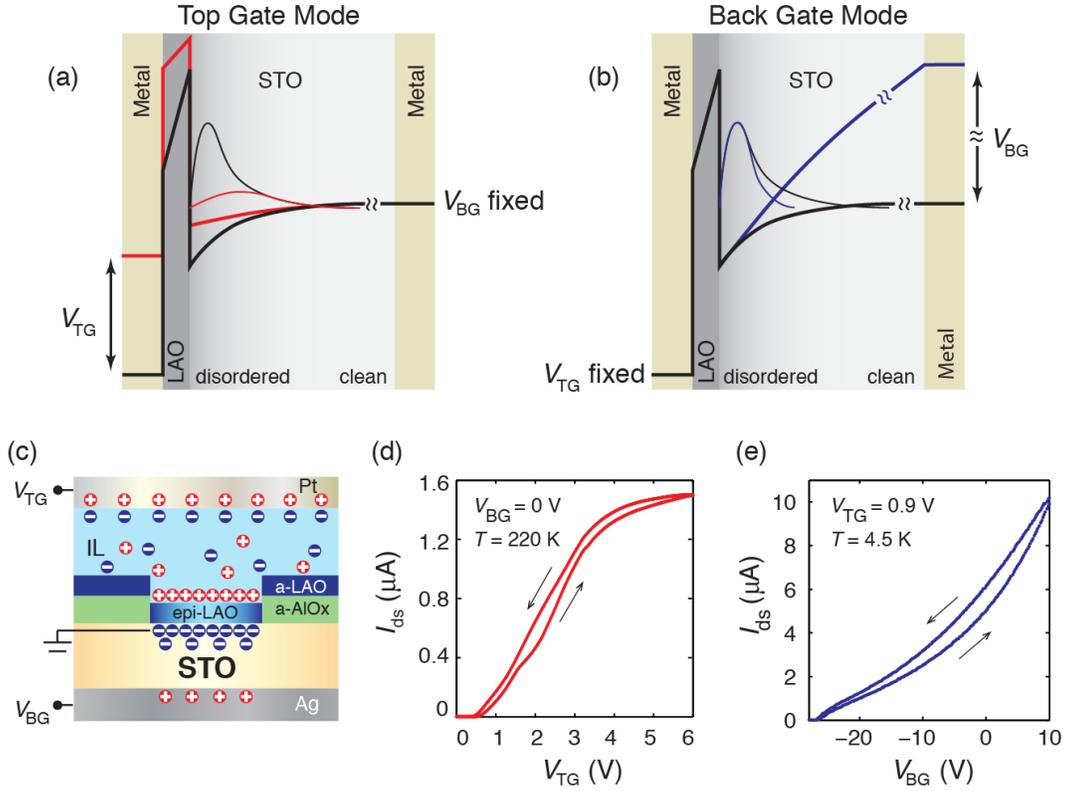

**Figure 1.** (**a**) Schematic diagram of a dual gated LaAlO$_3$/SrTiO$_3$ (LAO/STO) heterointerface system with varied $V_{TG}$ and zero $V_{BG}$. (**b**) The case of $V_{BG}$ modulation with fixed $V_{TG}$. Thick lines represent the conduction band bottom. Thin lines represent the electron envelope wavefunction. Note that in (a) and (b) the near interface region is magnified for clarity. (**c**) Schematic illustration of the hybrid liquid/solid dual gate LAO/STO device; "epi-" and "a-" indicate "epitaxial" and "amorphous", respectively. Amorphous AlO$_x$ is deposited prior to LAO growth for defining the Hall bar configuration. The central channel size is 250 μm × 50 μm. The epitaxial LAO layer thickness is 4 unit cell (1.5 nm). The STO substrate thickness is 0.5 mm. The Pt electrode (~ 3 × 5 mm$^2$) is about 2 mm above the channel. (**d**) Transfer curve for the ionic liquid top gate with fixed $V_{BG} = 0$ V at temperature $T = 220$ K just above the melting point of the ionic liquid. Drain-source voltage $V_{ds} = 0.1$ V. $I_{ds}$: drain-source current. (**e**) Transfer curve for the back gate with fixed $V_{TG} = 0.9$ V at $T = 4.5$ K. $V_{ds} = 0.01$ V.

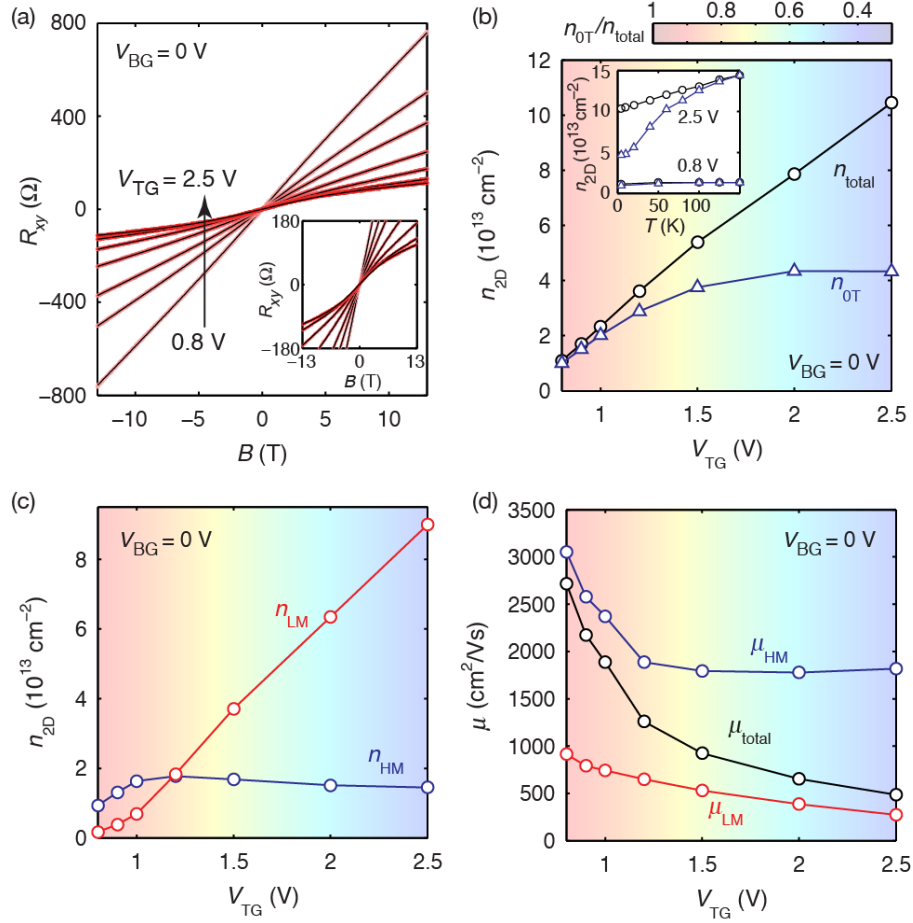

**Figure 2.** (a) Hall resistance as a function of magnetic field at $V_{TG}$ = 0.8 V, 0.9 V, 1.0 V, 1.2 V, 1.5 V, 2.0 V, 2.5 V. Measurement temperature $T$ = 2 K. Red curves are experiment data. Black curves are fits using the two-carrier model. Inset: magnification of the same data to emphasize the nonlinearity in the high-$V_{TG}$ regime. (b) Hall density as a function of $V_{TG}$. $n_{total}$ (black circles) and $n_{0T}$ (blue triangles) are Hall densities extracted using the slope $dR_{xy}/dB$ at 13 T and 0 T, respectively. The colour scale background (also in (c) and (d)) represents the ratio $n_{0T}/n_{total}$, which serves as a measure of the strength of the nonlinear Hall effect. Inset: temperature dependence of Hall densities $n_{total}$ and $n_{0T}$ for $V_{TG}$ = 2.5 V and $V_{TG}$ = 0.8 V. (c), (d) Two-carrier fitting results with $V_{TG}$ tuning. Red and blue circles represent low-mobility (LM) and high-mobility (HM) electrons, respectively. (c) Density of the two types of electrons $n_{LM}$ and $n_{HM}$ as a function of $V_{TG}$. (d) Mobility of the two types of electrons $\mu_{LM}$ and $\mu_{HM}$ as a function of $V_{TG}$. The black circles represent the total effective carrier mobility $\mu_{total} = \sigma/en_{total}$.

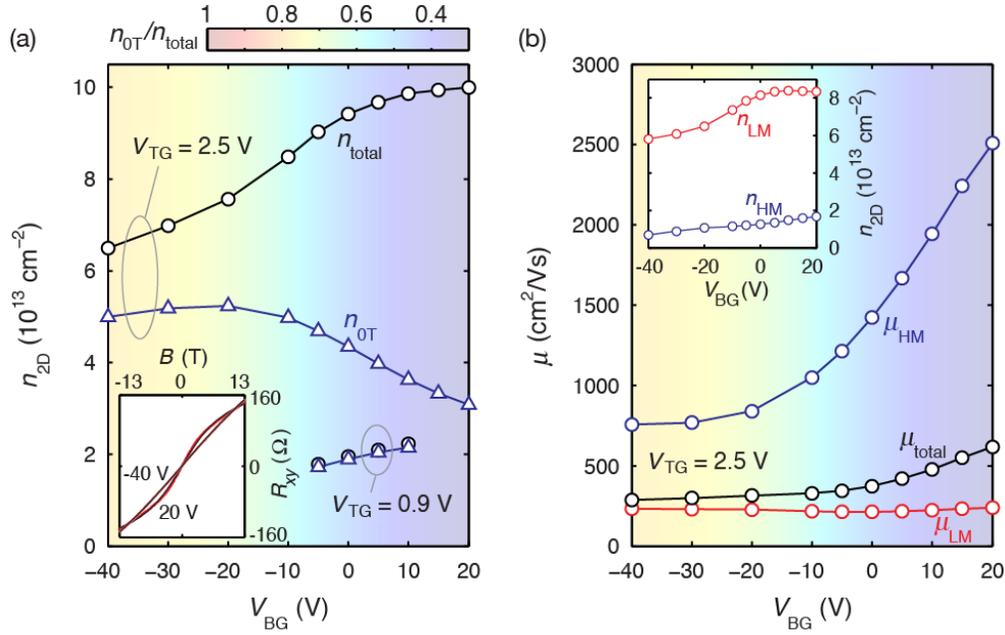

**Figure 3.** (**a**) Hall densities $n_\text{total}$ (black circles) and $n_\text{0T}$ (blue triangles) as a function of $V_\text{BG}$ at fixed $V_\text{TG} = 2.5$ V and $V_\text{TG} = 0.9$ V. Inset: Hall resistance as a function of magnetic field at two extreme back gate voltages ($V_\text{BG} = -40$ V and 20 V) at fixed $V_\text{TG} = 2.5$ V to show the suppression of nonlinearity at negative $V_\text{BG}$. Measurement temperature $T = 2$ K. (**b**) Two-carrier model fitting results with $V_\text{BG}$ tuning at fixed $V_\text{TG} = 2.5$ V. Red and blue circles represent low-mobility (LM) and high-mobility (HM) electrons, respectively. Main panel: mobility of the two types of electrons $\mu_\text{LM}$ and $\mu_\text{HM}$ as a function of $V_\text{BG}$, and the black circles represent total effective carrier mobility $\mu_\text{total}$. Inset: density of the two types of electrons $n_\text{LM}$ and $n_\text{HM}$ as a function of $V_\text{BG}$. The colour scale background represents the ratio $n_\text{0T}/n_\text{total}$ for the $V_\text{TG} = 2.5$ V case. Measurement temperature $T = 2$ K.

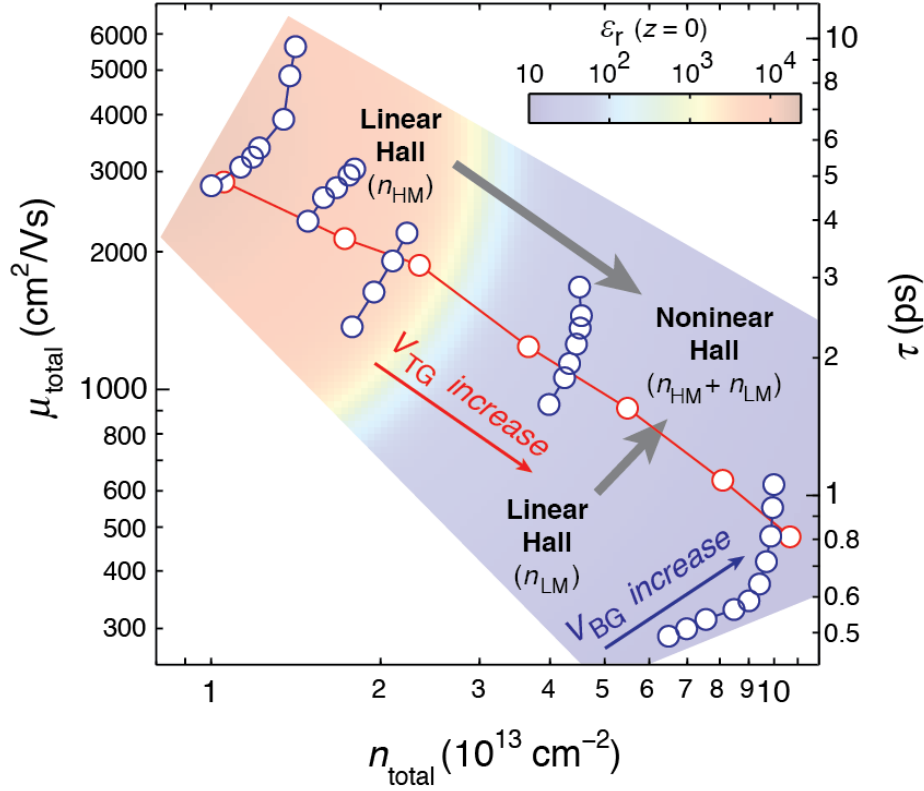

**Figure 4.** Bottom axis: total carrier density $n_{total}$. Left axis: total effective carrier mobility $\mu_{total}$. Right axis: estimated average scattering time of the total mobile electrons assuming a Drude model. Red circles correspond to top gate tuning data at $V_{BG} = 0$ V (same data as shown in Figure 2). Blue circles correspond to varied $V_{BG}$ for fixed $V_{TG} = 0.7$ V, 0.8 V, 0.9 V, 1.5 V, 2.5 V, in which $V_{TG} = 0.9$ V and 2.5 V cases are shown in Figure 3. Red and blue arrows indicate the direction of increasing top and back gate voltages, respectively. Background colour scale represents the estimated dielectric constant $\varepsilon_r$ of STO at the interface ($z = 0$), based on the Poisson-Schrödinger simulation. Red colour indicates near the full static value and the blue colour indicates suppressed values. All Hall effect and resistivity measurements shown were performed at temperature $T = 2$ K.

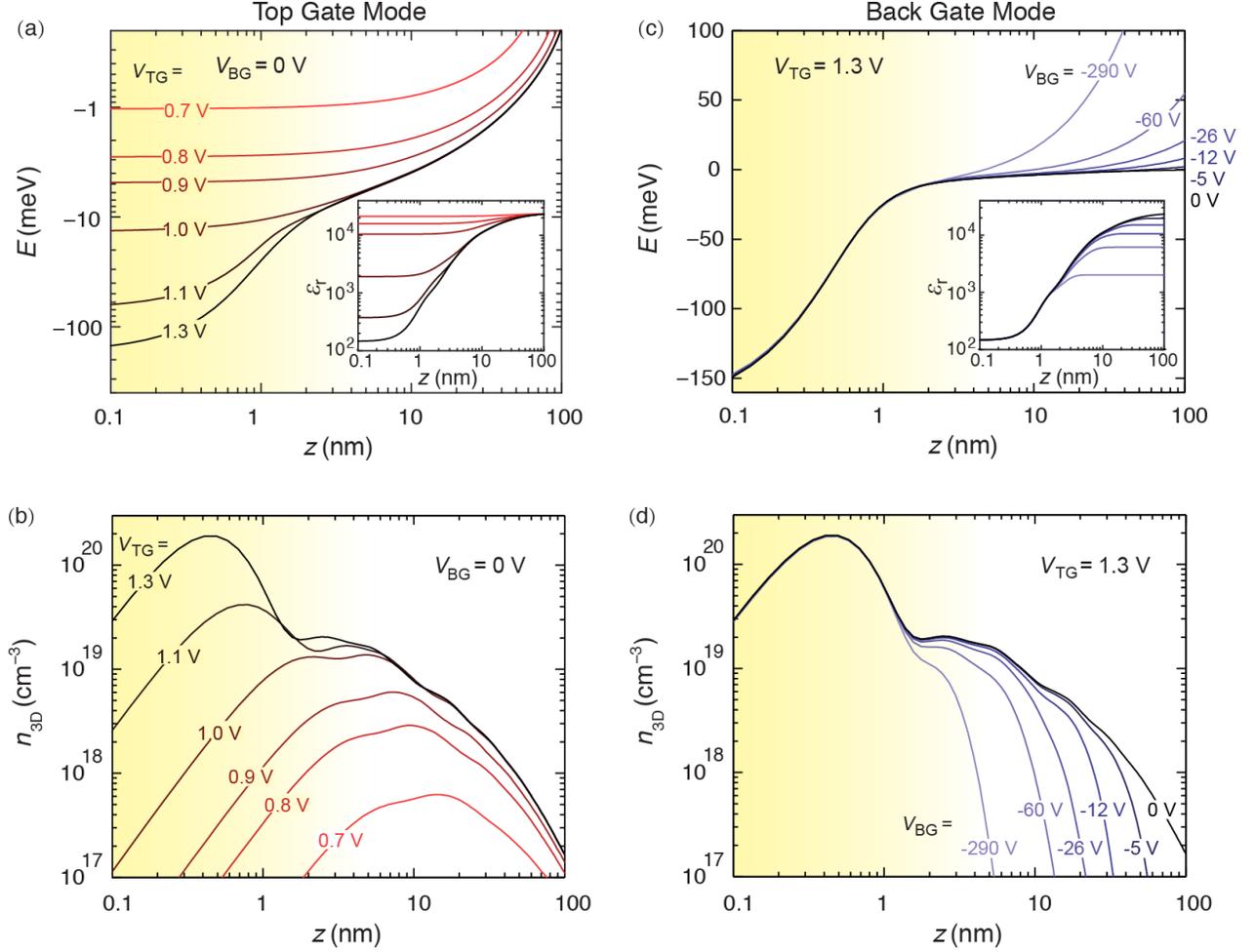

**Figure 5.** (**a**) The quantum well potential profile (the locally averaged effective potential in the continuum approximation) and (**b**) the spatial distribution of interface electrons in the depth direction $z$, under different $V_{TG}$ and zero $V_{BG}$ (top gate mode). $E$: conduction band bottom. $n_{3D}$: three-dimensional density of electrons. (**c**) and (**d**) show quantum well potential profiles and electron spatial distributions, respectively, under different $V_{BG}$ for fixed $V_{TG} = 1.3$ V (back gate mode). Inset of (a) and (c) show $\varepsilon_r$ of STO as a function of $z$ for top gate mode and back gate mode, respectively. Yellow shaded area highlights the near-interface region with more disorder. The size of this region can be related to the cation interdiffusion length scales at the interface[52]. All simulations are performed for $T = 2$ K.

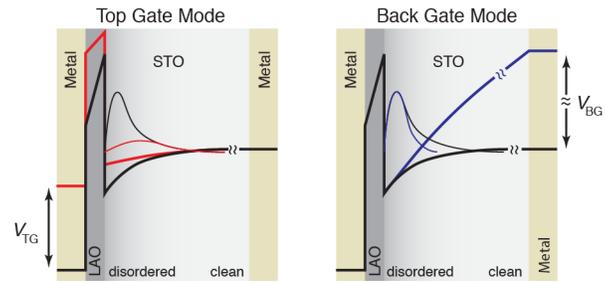

TOC Graphic